\documentclass[letterpaper,10pt]{article} 
\newcommand{\authormark}[1]{\textsuperscript{#1}}

\usepackage{osameet3} 

\usepackage{amsmath,amssymb}
\usepackage{upgreek}
\usepackage{multirow}
\usepackage{pgfplots}
\usepackage{graphicx}
\usepackage{subcaption}
\usepackage{mwe}
\usepackage{adjustbox}
\usepackage{xcolor}
\usepackage{collcell}
\usepackage{multirow}
\usepackage{mathtools}
\usepackage[colorlinks=true,bookmarks=false,citecolor=blue,urlcolor=blue]{hyperref} 

\usepackage{osameet3} 
\usepackage[font=small,skip=0pt]{caption}
\usepackage{amsmath,amssymb}
\usepackage[colorlinks=true,bookmarks=false,citecolor=blue,urlcolor=blue]{hyperref} 

\pgfplotsset{compat=1.17}
\begin{document}

\title{Power and Modulation Format Transfer Learning for Neural Network Equalizers in Coherent Optical Transmission Systems}

\vspace{-5mm}
\author{Pedro J. Freire \authormark{1}, Daniel Abode \authormark{1}, Jaroslaw E. Prilepsky \authormark{1}, Sergei K. Turitsyn \authormark{1}}
\address{\authormark{1} Aston Institute of Photonic Technologies, Aston University, B4 7ET, Birmingham, UK}

\copyrightyear{2021}

\vspace{-5mm}
\begin{abstract}
Transfer learning is proposed to adapt an NN-based nonlinear equalizer across different launch powers and modulation formats using a 450km TWC-fiber transmission. The result shows up to 92\% reduction in epochs or 90\% in the training dataset. 
\end{abstract}

\section{Introduction}\vspace{-1mm}
Neural networks (NN) have shown good performance in modeling nonlinear systems due to their flexible structure and the presence of nonlinear activation functions rendering the essential representation power\cite{Maximilian}. An optical transmission system can be seen as an example of a nonlinear system; the transfer functions of the fiber and components are nonlinear, causing noticeable distortion to the propagating signal, especially at high powers. Hence, it is desirable to reverse nonlinear effects to ensure signals' fidelity. There have been proposed some NN-based nonlinear equalizers, which has shown better accuracy in reversing the nonlinearities compared to conventional nonlinear Volterra and DBP-based equalization \cite{Maximilian, alan2020}. For NNs, to maintain a high enough performance, the training and test data set must be independent and identically distributed \cite{Tan}. However, realistic optical network parameters can be dynamic with changes in the power, modulation format, symbol rate, fiber distance, and the likes. Hence, the test data and the training data could be ``un-identically'' distributed, thereby flawing the NN application conditions. This flaw can cause a deterioration in the performance of the NN-based equalizer, as shown in the results section of this paper. One solution would be to train a new NN equalizer for each variation in channel and signal properties. However, this approach is rather impractical and computationally inefficient as training such an equalizer is data-hungry, requiring a large number of epochs and training time to reach optimal performance. We argue that a more efficacious approach can be to use transfer learning, where the initial NN setup is called the source, and the new NN variation -- the target \cite{Tan,Pan}. In transfer learning, some knowledge from the source model can be transferred to the target model, saving some noticeable percentage of both the training data and training epochs required to achieve optimal performance. Previous studies \cite{Xu,Jing} considered this approach for variations in bit rate, fiber length, and received optical power in a short reach direct detection optical network.
In this paper, we demonstrate the use of transfer learning for the adaption of an NN-based nonlinear equalizer designed for a source system of 16QAM 34.4~GBd 450~km TWC-fiber optical coherent system (the initial launch power is 5~dBm) to perform equalization in target systems (QAM-32, QAM-64, QAM-128) 34.4~GBd 450~km TWC-fiber optical coherent system operating at 2~dBm power. We consider the use of inductive network-based deep transfer learning \cite{Tan,Pan} which is more suitable for our problem of link functioning power alternation. The results show that with this method it is possible to realize up to 92\% reduction in epochs or 90\% reduction in the size of training data for the target model. This demonstrates the possibility of realizing a fast reconfigurable nonlinear equalizer, thereby edging us closer to a practical NN-based nonlinear equalizer for the next-generation optical network.

\section{Inductive NN-based transfer learning in nonlinear NN equalizers}\vspace{-1mm}
In our current work, we describe the transfer learning concept applied to the recently-introduced advanced equalizer which combines a convolutional layer with a bidirectional long-short term memory layer, see Ref.~\cite{pedro2021} for the detailed description of the equalizer and its performance. However, we argue that the proposed method is universal and can be applied to other NN equalizers. The input dataset is the distorted received signal with four features, the real and imaginary components of both polarizations. The labeled dataset is the true value of the transmitted symbol. In this study, we use the same NN topology, training methodology, and transmission simulator as in \cite{pedro2021}. We trained the source model for a system of 9 spans of 50~km TWC fiber, 16-QAM, 34.4~GBd optical network with a transmitter power of 5~dBm. We use inductive transfer learning to transfer some knowledge of the source model to the target model for 32-QAM, 64-QAM, and 128-QAM systems for the transmitter power of 2~dBm, while the other parameters are kept constant. The transfer process involves freezing some parameters of the source NN model and retraining some parameters to realize the target NN model. Let $W_s$, $b_s$, $W_T$, $b_T$ represent the spatial tensor of the weights and biases of the source NN model, $f(W_s,b_s)$, and of the target NN model, $f(W_T, b_T)$, respectively; the NNs are optimized by back-propagation using the ADAM optimizer. By inductive transfer learning,  $W_T = \{W_{s_f}, W_{s_t}\}$ and $b_T = \{b_{s_f}, b_{s_t}\}$ contain a trainable source NN parameters part ($W_{s_t}$ and  $b_{s_t}$) and  a fixed/frozen ones ($W_{s_f}$ and  $b_{s_f}$).  For this task, we found out that retraining the convolutional layer parameters while freezing the other layers' parameters provides the best performance.

\section{Results and Discussion}
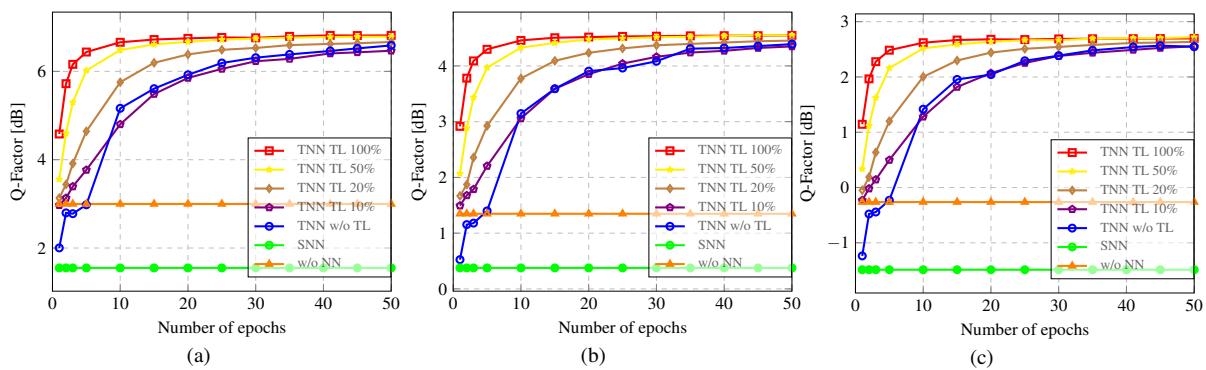
\begin{figure*}[ht]
	\begin{subfigure}{.32\textwidth}
  \begin{tikzpicture}[scale=0.65]
    \begin{axis} [
        xlabel={Number of epochs},
        ylabel={Q-Factor [dB]},
        grid=both,   
    	xmin=0, xmax=50,
        legend style={legend style={at={(1,0.3)},anchor= east}, legend cell align=left,fill=white, fill opacity=0.6, draw opacity=1,text opacity=1},
    	grid style={dashed}]
        ]
       \addplot[color=red, mark=square, very thick]
    coordinates {(1,4.5818)(2,5.7212)(3,6.1588)(5,6.4379)(10,6.6586)(15,6.7212)(20,6.7479)(25,6.7688)(30,6.7601)(35,6.7929)(41,6.8154)(45,6.8114)(50,6.815)};
    \addlegendentry{\footnotesize{TNN TL 100\%}};
    \addplot[color=yellow, mark=star, very thick]
   coordinates {(1,3.5542)(2,4.5966)(3,5.2976)(5,6.0205)(10,6.4825)(15,6.6113)(20,6.6682)(25,6.7184)(30,6.7422)(35,6.7641)(41,6.7769)(45,6.7828)(50,6.7838)};
    \addlegendentry{\footnotesize{TNN TL 50\%}};
    \addplot[color=brown, mark=diamond, very thick]
   coordinates {(1,3.1311)(2,3.4389)(3,3.9137)(5,4.6428)(10,5.7542)(15,6.1938)(20,6.3831)(25,6.4882)(30,6.5301)(35,6.5948)(41,6.6216)(45,6.6355)(50,6.666)};
    \addlegendentry{\footnotesize{TNN TL 20\%}};
    \addplot[color=violet, mark=pentagon, very thick]
   coordinates {(1,2.9697)(2,3.1314)(3,3.3928)(5,3.7679)(10,4.8009)(15,5.4869)(20,5.8512)(25,6.0563)(30,6.2293)(35,6.2827)(41,6.4056)(45,6.4311)(50,6.4671)};
    \addlegendentry{\footnotesize{TNN TL 10\%}};
    \addplot[color=blue, mark=o, very thick]   
    coordinates {(1,2.0021)(2,2.796)(3,2.7771)(5,2.9789)(10,5.1635)(15,5.6085)(20,5.9223)(25,6.1894)(30,6.3095)(35,6.3841)(41,6.4613)(45,6.5204)(50,6.5872)};
    \addlegendentry{\footnotesize{TNN w/o TL}};
    
    \addplot[color=green, mark=*, very thick]     
    coordinates {(1,1.549304)(2,1.549304)(3,1.549304)(5,1.549304)(10,1.549304)(15,1.549304)(20,1.549304)(25,1.549304)(30,1.549304)(35,1.549304)(40,1.549304)(45,1.549304)(50,1.549304)};
    \addlegendentry{\footnotesize{SNN}};

    \addplot[color=orange, mark=triangle, very thick]     
    coordinates {(1,2.9975)(2,2.9975)(3,2.9975)(5,2.9975)(10,2.9975)(15,2.9975)(20,2.9975)(25,2.9975)(30,2.9975)(35,2.9975)(41,2.9975)(45,2.9975)(50,2.9975)};
    \addlegendentry{\footnotesize{w/o NN}};
   \end{axis}
    \end{tikzpicture}
    \vspace*{-6mm} 
    \caption{ }
    \end{subfigure}~
    \begin{subfigure}{.31\textwidth}
  \begin{tikzpicture}[scale=0.65]
    \begin{axis} [
        xlabel={Number of epochs},
        ylabel={Q-Factor [dB]},
        grid=both,   
    	xmin=0, xmax=50,
        legend style={legend style={at={(1,0.3)},anchor= east}, legend cell align=left,fill=white, fill opacity=0.6, draw opacity=1,text opacity=1},
    	grid style={dashed}]
        ]
       \addplot[color=red, mark=square, very thick]
    coordinates {(1,2.91579)(2,3.78072)(3,4.09024)(5,4.29790)(10,4.45799)(15,4.50593)(20,4.51750)(25,4.529849)(30,4.53345)(35,4.540048)(40,4.54263)(45,4.54263)(50,4.54767)};
    \addlegendentry{\footnotesize{TNN TL 100\%}};
    \addplot[color=yellow, mark=star, very thick]
   coordinates {(1,2.0681)(2,2.8824)(3,3.4364)(5,3.9703)(10,4.3281)(15,4.4137)(20,4.4752)(25,4.4935)(30,4.5147)(35,4.5227)(40,4.5355)(45,4.548)(50,4.5486)};
    \addlegendentry{\footnotesize{TNN TL 50\%}};
    \addplot[color=brown, mark=diamond, very thick]
   coordinates {(1,1.669)(2,1.8716)(3,2.3572)(5,2.9249)(10,3.7758)(15,4.0918)(20,4.2351)(25,4.3145)(30,4.3704)(35,4.3947)(40,4.4196)(45,4.4429)(50,4.4539)};
    \addlegendentry{\footnotesize{TNN TL 20\%}};
    \addplot[color=violet, mark=pentagon, very thick]
   coordinates {(1,1.4939)(2,1.6778)(3,1.7873)(5,2.2045)(10,3.0579)(15,3.5897)(20,3.8516)(25,4.0412)(30,4.163)(35,4.2393)(40,4.2708)(45,4.322)(50,4.35)};
    \addlegendentry{\footnotesize{TNN TL 10\%}};
    \addplot[color=blue, mark=o, very thick]   
    coordinates {(1,0.52491)(2,1.1535)(3,1.1803)(5,1.3992)(10,3.1461)(15,3.5907)(20,3.9064)(25,3.9619)(30,4.0834)(35,4.3097)(40,4.3212)(45,4.356)(50,4.3905)};
    \addlegendentry{\footnotesize{TNN w/o TL}};
    
    \addplot[color=green, mark=*, very thick]     
    coordinates {(1,0.37647)(2,0.37647)(3,0.37647)(5,0.37647)(10,0.37647)(15,0.37647)(20,0.37647)(25,0.37647)(30,0.37647)(35,0.37647)(40,0.37647)(45,0.37647)(50,0.37647)};
    \addlegendentry{\footnotesize{SNN}};

    \addplot[color=orange, mark=triangle, very thick]     
    coordinates {(1,1.3479)(2,1.3479)(3,1.3479)(5,1.3479)(10,1.3479)(15,1.3479)(20,1.3479)(25,1.3479)(30,1.3479)(35,1.3479)(40,1.3479)(45,1.3479)(50,1.3479)};
    \addlegendentry{\footnotesize{w/o NN}};
   \end{axis}
    \end{tikzpicture}
    \vspace*{-6mm} 
    \caption{ }
    \end{subfigure}~
        \begin{subfigure}{.31\textwidth}
        \begin{tikzpicture}[scale=0.65]
        \begin{axis} [
        xlabel={Number of epochs},
        ylabel={Q-Factor [dB]},
        grid=both,   
    	xmin=0, xmax=50,
        legend style={legend style={at={(1,0.3)},anchor= east}, legend cell align=left,fill=white, fill opacity=0.6, draw opacity=1,text opacity=1},
    	grid style={dashed}]
        ]
      \addplot[color=red, mark=square, very thick]
    coordinates {(1,1.1434)(2,1.9654)(3,2.2763)(5,2.4841)(10,2.6199)(15,2.6683)(20,2.6811)(25,2.6763)(30,2.6883)(35,2.6909)(41,2.6926)(45,2.6952)(50,2.6968)};
    \addlegendentry{\footnotesize{TNN TL 100\%}};
    \addplot[color=yellow, mark=star, very thick]
   coordinates {(1,0.33518)(2,1.117)(3,1.6244)(5,2.1611)(10,2.518)(15,2.5883)(20,2.6354)(25,2.6644)(30,2.6644)(35,2.6922)(41,2.7002)(45,2.704)(50,2.7199)};
    \addlegendentry{\footnotesize{TNN TL 50\%}};
    \addplot[color=brown, mark=diamond, very thick]
   coordinates {(1,-0.049313)(2,0.18723)(3,0.63557)(5,1.2001)(10,2.0021)(15,2.2971)(20,2.4407)(25,2.5092)(30,2.5435)(35,2.579)(41,2.6159)(45,2.6213)(50,2.6345)};
    \addlegendentry{\footnotesize{TNN TL 20\%}};
    \addplot[color=violet, mark=pentagon, very thick]
   coordinates {(1,-0.23071)(2,-0.018991)(3,0.143)(5,0.49762)(10,1.2809)(15,1.8189)(20,2.0695)(25,2.2527)(30,2.3793)(35,2.4373)(41,2.4904)(45,2.5246)(50,2.5541)};
    \addlegendentry{\footnotesize{TNN TL 10\%}};
    \addplot[color=blue, mark=o, very thick]   
    coordinates {(1,-1.2374)(2,-0.47904)(3,-0.44315)(5,-0.2298)(10,1.4166)(15,1.9549)(20,2.0419)(25,2.2927)(30,2.3859)(35,2.4812)(41,2.5405)(45,2.5658)(50,2.5474)};
    \addlegendentry{\footnotesize{TNN w/o TL}};
    
    \addplot[color=green, mark=*, very thick]     
    coordinates {(1,-1.4870)(2,-1.4870)(3,-1.4870)(5,-1.4870)(10,-1.4870)(15,-1.4870)(20,-1.4870)(25,-1.4870)(30,-1.4870)(35,-1.4870)(40,-1.4870)(45,-1.4870)(50,-1.4870)};
    \addlegendentry{\footnotesize{SNN}};

    \addplot[color=orange, mark=triangle, very thick]     
    coordinates {(1,-0.26304)(2,-0.26304)(3,-0.26304)(5,-0.26304)(10,-0.26304)(15,-0.26304)(20,-0.26304)(25,-0.26304)(30,-0.26304)(35,-0.26304)(41,-0.26304)(45,-0.26304)(50,-0.26304)};
    \addlegendentry{\footnotesize{w/o NN}};
   \end{axis}
    \end{tikzpicture}
    \vspace*{-6mm} 
    \caption{ }
    \end{subfigure}~
    \caption{Transfer learning from Source System 5~dBm TWC 9$\times$50~km 16-QAM 34.4 GBd to target system with 2 dBm power and (a) 32-QAM, (b) 64-QAM (c) 128-QAM 34.4 GBd TWC 9$\times$50~km trained with different size of training data: 100\%, 50\%, 20\%, and 10\%.}
    \label{Ideal_to_NotIdea2QAMpercent}
  \end{figure*}

Fig.~\ref{Ideal_to_NotIdea2QAMpercent} compares the performance in terms of Q-factor for the target NN (TNN) models realized through transfer learning with different sizes of training data (``TNN x\%'' label) to TNN trained from random initialization without transfer learning (``TNN w/o TL''). We also show there the source NN when tested with target data (``SNN'') as well as the Q-factor without a NN equalizer (``w/o NN''). Figs.~\ref{Ideal_to_NotIdea2QAMpercent} (a), (b) and (c) depict the results for the different modulation formats: 32-QAM, 64-QAM, and 128-QAM target systems, respectively. By comparing the plots in each panel, we observe that the least Q-factor occurred when the SNN was tested with data from the target system. This curve lies even below the reference orange line corresponding to the non-equalized case. It justifies the hypothesis that the NN equalizer performance would deteriorate when tested on new datasets from an ``un-identical'' distribution, in this case, the target systems of different modulation formats and transmitter power. The plot shows that using transfer learning (labeled with TL), we can achieve TNN performance similar to TNN w/o TL with as little as 10\% of the needed training data (shown as TNN 10\% ), which represents a reduction in the training dataset from 262k symbols to 26k symbols. Furthermore, by training with TL, the result of TNN 100\% compared to TNN w/o TL indicates that we can achieve the best Q-factor with just 4 epochs in contrast to 50 epochs when we train the NN from scratch. It justifies that transfer learning can help in the fast adaptation of a nonlinear NN equalizer to changes in modulation format with the possibility of about 92\% reduction in epoch or 90\% reduction in training data, which also implies a considerable reduction in training time since we are retraining just one layer of the equalizer.

\vspace*{-9mm}  
\section{Conclusion}
We demonstrated the advantage of transfer learning in adapting an NN nonlinear equalizer trained for a source system of 16-QAM and 5~dBm power to target systems of different modulation formats and 2~dBm power. The result shows that transfer learning can help reduce the epochs and size of data required. Hence, an expedited reconfiguration for changes in the transmission system can be achieved.
\vspace*{-2mm}  
\section{Acknowledgements}
\footnotesize
The EU Horizon 2020 program under the Marie Sklodowska-Curie grant agreement No. 813144 (REAL-NET); SMARTNET  EMJMD  program  (Project number 586686-EPP-1-2017-1-UK-EPPKA1-JMD-MOB); EPSRC project TRANSNET; Leverhulme project RPG-2018-063.


\begin{thebibliography}{99} 

\footnotesize


\bibitem{Maximilian} M. Schaedler, et al ``Deep Neural Network Equalization for Optical Short Reach Communication,''Appl. Sci.  \textbf{9}, 4675 (2019).

\bibitem{alan2020},
Q. Fan, et al.,  ``Advancing theoretical understanding and practical performance of signal processing for nonlinear optical communications through machine learning'', Nat. Commun.,  vol. 11, art. no. 3694, 2020.


\bibitem{Tan} C. Tan, F. Sun, T. Kong, W. Zhang, C. Yang, and C. Liu. “A Survey on Deep Transfer Learning,” ICANN, pp. 270--279, 2018.
\bibitem{Pan}  S. J. Pan and Q. Yang, "A Survey on Transfer Learning," in IEEE Transactions on Knowledge and Data Engineering, vol. 22, pp. 1345-1359, 2010, 
\bibitem{Xu} Z. Xu, C. Sun, T. Ji, H. Ji, and W. Shieh, "Transfer Learning Aided Neural Networks for Nonlinear Equalization in Short-Reach Direct Detection Systems," OFC 2020, paper T4D.4.
\bibitem{Jing} J. Zhang, et al.  ''Fast remodeling for nonlinear distortion mitigation based on transfer learning,'' Opt. Lett. 44, 4243-4246 (2019).
\bibitem{pedro2021} P. J. Freire et al., ''Performance versus Complexity Study of Neural Network Equalizers in Coherent Optical Systems,'' preprint arXiv:2103.08212, 2021.


\end{thebibliography}
\end{document}